\documentclass[%
aps,prl,twocolumn,
superscriptaddress,
]{revtex4-2}

\usepackage{graphicx}
\usepackage{amsfonts}
\usepackage{amssymb}
\usepackage{amsthm}
\usepackage{amsmath}
\usepackage{multirow}

\usepackage{color}
 \usepackage{makecell}
\usepackage{mathtools}

\usepackage[thinc]{esdiff}
\usepackage[colorlinks=true,allcolors=blue]{hyperref}

\begin{document}
	
\title{Monopoles, Strings, Walls and Gravitational waves}
\author{Rinku Maji}
\affiliation{Cosmology, Gravity and Astroparticle Physics Group, Center for Theoretical Physics of the Universe,  Institute for Basic Science, Daejeon 34126, Republic of Korea}
	\author{Qaisar Shafi}
		\affiliation{Bartol Research Institute, Department of Physics and 
		Astronomy,
		 University of Delaware, Newark, DE 19716, USA}
\begin{abstract}
The gauge symmetry breaking $SU(2) \to U(1) \to Z_2 \to 1$ successively produces monopoles, strings and domain walls bounded by strings (WBS), with the elementary monopole carrying a $U(1)$ magnetic flux twice as large as the elementary string. The elementary strings and subsequently WBS emit gravitational waves, and during their decay, the WBS yield a network of composite strings that carry the same $U(1)$ flux as the monopoles. Depending on the cosmological evolution, we provide the gravitational wave spectra generated by the elementary strings and WBS, in combination with the composite strings which are either effectively stable, quasistable, or metastable. These three scenarios are also realized in the symmetry breaking chain $SU(3) \to SO(3) \to Z_2 \to 1$. Both $SU(2)$ and $SU(3)$ have appeared in the literature as flavor gauge symmetries.

\end{abstract}

\maketitle
%%%%%%%%%%%%%%%%%%%%%%%%%%%%%%%%%%%%%%%%%%%%%%%%%%%%%%%
\section{Introduction}
%%%%%%%%%%%%%%%%%%
Grand unified theories based on gauge groups  $SO(10)$, more precisely $Spin(10)$ \cite{Georgi:1974my,Fritzsch:1974nn}, and $E_6$ \cite{Gursey:1975ki,Achiman:1978vg,Shafi:1978gg}, predict a variety of topological structures, including magnetic monopoles \cite{tHooft:1974kcl,Polyakov:1974ek, Lazarides:1980cc} and cosmic strings \cite{Kibble:1982ae}. A particularly well known topological structure is a domain wall bounded by strings (WBS), first discovered \cite{Kibble:1982dd} 
 in $SO(10)$, and subsequently in a variety of condensed matter systems
including superfluid He$^3$-B \cite{Makinen:2018ltj}. For a comprehensive discussion of topological defects in $SO(10)$, see Refs.~\cite{Lazarides:2019xai, Lazarides:2023iim, Maji:2025thf}.

Following the discovery of gravitational waves \cite{LIGOScientific:2016aoc}, there has been a great deal of interest in cosmic strings and WBS because the gravitational emission by these structures can be tested by a variety of ongoing and proposed experiments. Metastable strings \cite{Buchmuller:2021mbb, NANOGrav:2023hvm}, quasistable strings \cite{Lazarides:2022jgr, Lazarides:2023ksx} and WBS \cite{Maji:2023fba}, for instance, with a dimensionless string tension $G \mu$ of order $10^{-6} -10^{-7}$, have been shown to yield a spectrum which is compatible with the stochastic gravitational wave background (SGWB) reported by the pulsar timing array (PTA) experiments \cite{NANOGrav:2023gor, EPTA:2023fyk, Reardon:2023gzh, Xu:2023wog}. 

Recent studies have also focused on the gravitational emission from WBS in $SO(10)$ \cite{Dunsky:2021tih, Maji:2023fba, Maji:2025yms}, from $U(1)$ flavor gauge symmetry \cite{Ghoshal:2026wwu} which contains WBS, and from metastable strings appearing in $SU(2)$ flavor gauge symmetry \cite{Antusch:2025xrs} (for other related studies, see Refs.~\cite{Buchmuller:2021dtt, Buchmuller:2023aus, Lazarides:2023rqf, Maji:2023fhv, Afzal:2023kqs, Afzal:2023cyp, Fu:2023mdu, Chitose:2023dam, Bao:2024bws, Maji:2024tzg, Maji:2024cwv, Datta:2024bqp, Chitose:2025qyt, Datta:2025owx, Ingoldby:2025wcl, Tranchedone:2026lav, Asl:2026zpj, Blasi:2026iyq, deGiorgi:2026fyx, Hua:2026mgn}). In this paper we explore the cosmological consequences of $SU(2)$ gauge symmetry breaking to the identity element via $U(1)$ and $Z_2$. This breaking yields monopoles, strings and domain walls, and composite structures including monopole-antimonopole dumbbells, WBS, and wall bounded by a necklace \cite{Lazarides:2023iim}. Depending on the assumptions related to inflation and quantum tunneling, we can realize in this framework the metastable string scenario, or the quasistable string scenario, or a scenario with effectively stable strings. This is made possible during the collapse of WBS which yields a network of composite strings that carry the same flux as the $U(1)$ monopole. For the effectively stable string scenario, we estimate the gravitational wave spectrum for a dimensionless string tension $G \mu = 10^{-10}$ and $G\mu = 10^{-12}$, and with a $Z_2$ symmetry breaking scale of $10^2$ and $10^5$ GeV.  Our calculations show that the predicted gravitational-wave spectrum can be tested in a variety of ongoing and proposed experiments. We also investigate the quasistable and metastable string scenarios, and for $G\mu\sim 10^{-7}$ and $v_{\rm DW}\sim10^8\,\mathrm{GeV}$, the delayed formation of the string network following the collapse of WBS yields for both cases a spectrum that is compatible with the current LIGO-VIRGO-KAGRA (LVK) Run 4 bound \cite{LIGOScientific:2025bgj, LIGOScientific:2025kry} while explaining the pulsar timing array data.
We briefly show that these three scenarios can be realized with a gauge $SU(3)$ symmetry breaking chain 
$SU(3) \to SO(3) \to Z_{2} \to 1$.
%%%%%%%%%%%%%%%%%%%%%%%%%%%%%%%%%%%%%%%%%%%%%%%%%%%%%%%%%%%%%%%%%%
\section{Composite topological structures in $SU(2)$ and $SU(3)$}
\label{sec:2}
%%%%%%%%%%%%%%%%%%
Consider the gauge symmetry breaking
\begin{align}
\label{eq:sym-brkngZ2}
SU(2)\to Z_2\to 1.
\end{align}
Since $\pi_1(SU(2)/Z_2)=Z_2\equiv \lbrace +1, -1\rbrace$, the  first breaking $SU(2)\to Z_2$ produces $Z_2$-strings. Therefore, a string with even winding is topologically trivial. The WBS produced from the breaking of $Z_2\to 1$ connects the two $Z_2$-strings which annihilate. The WBS decay leaves behind no  topologically stable defect, consistent with the fact that $SU(2)$ is finally broken to the identity element.

 Next, consider the gauge symmetry breaking chain
\begin{align}
\label{eq:sym-brkng1}
SU(2)\to U(1)\to Z_2\to 1 .
\end{align}
The first breaking $SU(2)\to U(1)$ produces  a 't Hooft-Polyakov monopole \cite{tHooft:1974kcl,Polyakov:1974ek} that carries two units of $U(1)$ magnetic flux ($4 \pi / g$), where $g$ denotes the $SU(2)$ gauge coupling.  The flux corresponds to a $4\pi$ rotation around the $SU(2)$ generator $T_3=\mathrm{diag}[1,-1]$ associated with the unbroken $U(1)$. The second breaking $U(1)\to Z_2$ produces strings and the monopole flux is squeezed into two confined flux tubes, each carrying one unit of flux $2\pi/g$, associated with a $2\pi$ rotation around  $T_3$. The topological charge associated with  the string is given by 
\begin{align}
\label{eq:winding1}
\pi_1(U(1)/Z_2)=Z \equiv \lbrace 0, \pm 1, \pm 2, \pm 3, ... \rbrace.
\end{align}
 The monopole-antimonopole pairs on the string can form cosmic necklaces \cite{Lazarides:2019xai, Lazarides:2023iim}.

\begin{figure}[h!]
\begin{center}
\includegraphics[scale=0.5]{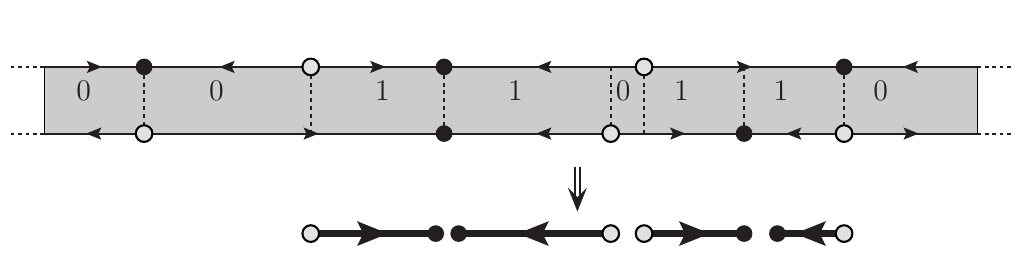}
\caption{Collapse of a domain wall bounded by cosmic strings with monopole-antimonopole pairs on them. The  anti-parallel strings annihilate since the fluxes are in opposite directions, as depicted by regions with zero topological charge.
 On the other hand, a wall bounded by parallel strings each having a unit topological flux collapses to form a composite string carrying the same flux as the magnetic monopole.}
\label{fig:WBS_QSS}
\end{center}
\end{figure}

Finally, the breaking of $Z_2$ generates domain walls attached to the strings that carry one unit of magnetic flux. These walls bounded  by strings, with monopole-antimonopole pairs on them, will collapse and yield monopole-antimonopole pairs connected by composite strings that now carry two units $(4\pi/g)$ of $U(1)$ magnetic flux, as shown in Fig.~\ref{fig:WBS_QSS}. The annihilation proceeds as follows.
%%%%%%%%%%%%%%%%%%%%%%%%%%%%%
%%%%%%%%%%%%%%%%%%%%%%%%%%%%%%%%%%%%%%%%%%%%%%%%%%%%%%%%%%%
\begin{itemize}
\item  A wall bounded by anti-parallel strings with winding numbers $\pm 1$ disappears since the net flux is zero.
\item  On the other hand, the collapse of a wall bounded by parallel strings yields a string carrying the same flux ($4\pi/g$) as the minimal monopole.
\end{itemize}

  We expect to find walls bounded by parallel or anti-parallel strings  even if the primordial magnetic monopoles are inflated away. A fraction ($f_s$) of these composite structures associated with parallel minimal strings should lead to the formation of doubly charged strings, while the rest $(1-f_s)$, with anti-parallel strings will disappear. Naively one may expect $f_s$ to be around $0.5$ or so. These strings are not topologically stable since quantum tunneling of monopole-antimonopole pairs will make them unstable. However, these doubly charged strings are effectively stable if their decay lifetime is equal to or greater than the age of the Universe.
  %Note that, if we start with a $U(1)$ gauge symmetry and follow the symmetry breaking pattern $U(1)\to Z_2\to I$, a fraction $f_s$ of the walls bounded by strings are expected to merge into a topologically stable cosmic string.

  %%%%%%%%%%%%%%%%%%%%%%%%%%%%%%%%%%%%%%%%%%%%%%%%%%%%%%%%%%%
  The topological structures considered here radiate gravitational waves, and  the gravitational radiation depends, among other things, on the string tension $\mu$ and the wall tension $\sigma\sim v_{\rm DW}^3$, where $v_{\rm DW}$ represents the vacuum-expectation-value (VEV) associated with the domain wall formation. Before cosmic time $R_c=\mu/\sigma$, the string dynamics dominates the evolution of WBS. The minimal strings are assumed to form a scaling network \cite{Vachaspati:1984gt,Vilenkin:2000jqa} with a characteristic length $\xi(t)\sim t$ and gravitational waves are emitted in the usual way. After $R_c$, the wall dynamics dominates and the WBS network, with typical scale $R_c$, finally collapses at time $t_d=1/(G\sigma)$ \cite{Hiramatsu:2013qaa, Dunsky:2021tih}. The collapsing WBS emit gravitational waves which is accompanied by the appearance of composite strings that carry the same flux as the elementary monopoles. 
  Following the collapse of WBS structures, there can be several remnant sources of gravitational waves, depending on the symmetry breaking pattern and cosmological evolution.
\begin{itemize}
\item For the breaking pattern \eqref{eq:sym-brkngZ2}, the network disappears and there is no source of gravitational waves.
\item For the breaking pattern \eqref{eq:sym-brkng1}, we could have the following scenarios  depending on the characteristic separation between the primordial monopole and antimonopole on the fundamental string:
\begin{enumerate}
\item  If the separation between the monopole and antimonopole on the fundamental strings remains of the order of the particle horizon, the WBS network collapses into subhorizon monopole-antimonopole pairs connected by composite strings. These structures radiate gravitational waves \cite{Martin:1996cp, Dvali:2022vwh} and eventually decay. At this point it is worth clarifying that $R_c$ provides the onset for the merger of WBS network, whereas the characteristic separation between the primordial monopole and antimonopole on the fundamental strings provides the length scale of the connected composite strings.
\item   In the absence of primordial monopoles, the WBS network merges into a connected  network of long composite strings  carrying two units ($4\pi/g$) of flux. This can happen, for example, when the primordial monopoles are inflated away. These strings could be metastable or effectively stable depending on whether the lifetime of decay via quantum-mechanical tunneling is shorter or longer than the age of the Universe.
\item  A network of quasistable strings appears if the primordial monopoles are only partially inflated and reenter the horizon at a later time $t_M$. 
\end{enumerate}
%\item For the symmetry breaking pattern $U(1)\to Z_2\to I$, a fraction $f_s$ of the walls bounded by strings can merge into the topologically stable cosmic strings.
\end{itemize}

Before concluding this section, let us briefly show that the topological structures appearing from $SU(2)$ breaking \eqref{eq:sym-brkng1} are in one-to-one correspondence with the following $SU(3)$ gauge symmetry breaking
\begin{align}
\label{eq:sym-brkng-SU3}
SU(3) \to SO(3) \to Z_{2R} \to 1.
\end{align}
%The topological structures appearing here, it turns out, are in one to one correspondence with the SU(2) model we have already discussed. 
The first breaking produces a $Z_2$ monopole since the second homotopic group
$\pi_2 (SU(3) / SO(3)) = \pi_1(SO(3))
= Z_2.$
The second breaking produces a $Z_4$ string which follows from the first homotopic group $\pi_1 (SO(3)/Z_{2R})$, where $Z_{2R} = \lbrace 1, R\rbrace$, and $R$ denotes a rotation by $\pi$ radians in $SO(3)$.
Since $SO(3) = SU(2)/Z_2$, where $Z_2 =\lbrace 1,-1\rbrace$ denotes the center of $SU(2)$, the coset space $SO(3)/Z_{2R}$ is homeomorphic to $SU(2)/Z_4$ by the third isomorphism theorem. Hence the breaking of $SO(3)$ to $Z_{2R}$ produces $Z_4$ strings. As a consequence, the $Z_2$ monopole flux is squeezed inside two separate tubes.
Finally, the breaking of $Z_{2R}$ in Eq.~\eqref{eq:sym-brkng-SU3} to the identity element produces a domain wall and, similar to the $SU(2)$ case discussed earlier, this yields a variety of topological structures including wall bounded by a necklace with the monopoles playing the role of beads on the string.
%%%%%%%%%%%%%%%%%%%%%%%%%%%%%%%%%%%%%%%%%%%%%%%%%%%%%%%%%%%  
\section{Gravitational wave backgrounds}
\label{sec:GWs}
%%%%%%%%%%%%%%%%%%%%%%%%%%%%%%%%%%%%%%%%%%%%%%%%%%%%%%%%%%% 
In this section we estimate the gravitational wave backgrounds from effectively stable strings, from quasistable strings, and from the metastable strings.
 The radiation from the string loops produced at  times $t_i\in\left[t_{\rm ini},t_{\rm fin}\right]$ in the scaling regime \cite{Vachaspati:1984gt,Vilenkin:2000jqa} can be expressed as
\begin{align}
\Omega_{\rm GW}^{\rm str}(f,t_{\rm ini},t_{\rm fin}) = & \frac{1}{\rho_{c,0}} \sum_{k=1}^{\infty}  \int_{t_{\rm ini}}^{t_0} d\tilde{t} \left(\frac{a(\tilde{t})}{a(t_0)}\right)^5 \nonumber \\ & \frac{\mathcal{F} C_{\rm eff}(t_i)}{(\Gamma G \mu + \alpha)\alpha t_i^4} \left(\frac{a(t_i)}{a(\tilde{t})}\right)^3 \frac{\Gamma k^{-n}}{\zeta(n)} \nonumber \\ &  G\mu^2 \frac{2 k}{f}\Theta(t_{\rm fin} - t_i) \Theta(t_i - t_{\rm ini}).
\end{align}
Here, $\rho_{c,0}$ is the present day ($t_0$) critical energy density, the numerical factors $\mathcal{F}\simeq 0.1$,  $C_{\rm eff} = 5.7$ in the radiation dominated era \cite{Vanchurin:2005pa,Ringeval:2005kr,Olum:2006ix,Olmez:2010bi,Blanco-Pillado:2013qja,Blanco-Pillado:2017oxo,Cui:2018rwi}, and the present day frequency of the gravitational waves in a normal mode $k$ radiated at time $\tilde{t}$ is given by
\begin{align}
f = \frac{a(\tilde{t})}{a(t_0)}\frac{2k}{\alpha t_i -\Gamma G\mu (\tilde{t}-t_i)} , \quad k=\lbrace 1,2,3,...\rbrace,
\end{align}
 where $a(t)$ denotes the scale factor, and the numerical factors $\Gamma \simeq 50$ and $\alpha\simeq 0.1$ \cite{Vachaspati:1984gt,Vilenkin:2000jqa}. We assume $n=4/3$ for the dominated gravitational radiation from the bursts from the cusp \cite{Damour:2001bk}.

The string dynamics dominates until time scale $R_c=\mu/\sigma$ \cite{Martin:1996ea}, after which the wall dynamics dominates, WBS structures oscillate and finally collapse at time  $t_d=1/(G\sigma)$ \cite{Hiramatsu:2013qaa, Dunsky:2021tih} radiating gravitational waves. The gravitational wave background from the WBS structures is expressed as
 \cite{Maji:2025yms}
\begin{align}
\Omega_{\rm GW}^{\rm dw}(f)=\mathcal{G}(t_d)\Omega_{r,0}\tilde{\Omega}^{\rm DW}_{\rm GW}(\tilde{f},t_d),
\end{align}
where $\mathcal{G}(t)=\tfrac{g_*(t)}{g_*(t_0)}\tfrac{g_{*s}(t_0)^{4/3}}{g_{*s}(t)^{4/3}}$, $\Omega_{r,0}=9.1476\times 10^{-5}$ \cite{Planck:2018vyg} and 
\begin{align}
\tilde{\Omega}^{\rm DW}_{\rm GW}(\tilde{f},t_d) = 
\begin{cases}
 \frac{\sigma w}{m_{\rm Pl}^2} (\tilde{f} w)^2 \quad \mathrm{for} \ \frac{1}{w}\frac{a(R_c)}{a(t_d)}\leq \tilde{f} \leq \frac{1}{w} \\
 \frac{\sigma w}{m_{\rm Pl}^2} (\tilde{f} w)^{-1}, \quad \mathrm{for} \  \tilde{f} \geq \frac{1}{w} \ 
\end{cases},
\end{align}
with $w\sim R_c$ and $\tilde{f}=\frac{a(t_0)}{a(t_d)}f$.
After $R_c$, rings of size $\alpha_r t$ ($\alpha_r$ is an $\mathcal{O}(1)$ numerical factor) can form \cite{Bao:2024bws} at time $t$ from the reconnection of the WBS structures. A fraction of these rings becomes loops after a time $t_d$, where $t_d = (G\mu)^{-1}R_c$. The emergent strings reenter the scaling regime after the collapse of WBS structures, and we parametrize it by a delay factor $d_s$ defined as 
\begin{align}
\label{eq:ds}
t_s = d_s R_c ,
\end{align}
such that the scaling distribution of loops $n(l,t\gtrsim t_d)$ starts from time $t_s$. Therefore, for $d_s\gtrsim (G\mu)^{-1}$, the loops are sourced by the long composite strings generated from the collapsing WBS.

 The gravitational wave background for the effectively stable and quasistable strings can be expressed as
\begin{align}
\Omega_{\rm GW}^{\rm str}(f,t_F,R_c) + \Omega_{\rm GW}^{\rm dw}(f)+ \Omega_{\rm GW}^{\rm str}(f,t_s,t_{\rm end}),
\end{align}
where $t_F={\rm max}[\tfrac{t_{\rm Pl}}{(G\mu)^2},\tfrac{t_{\rm Pl}}{\Gamma^2 (G\mu)^{5/2}}]$ is determined by the maximum timescale between friction dominated era for the primordial string network \cite{Vilenkin:1991zk, Garriga:1993gj} and the particle radiation domination from the string loops \cite{Blanco-Pillado:1998tyu,Matsunami:2019fss, Auclair:2019jip}. We take $t_{\rm end} = t_0$ for the effectively stable cosmic strings. In the case of quasistable strings, we set $t_{\rm end}=t_M$, which is the horizon reentry time for the partially inflated monopoles. 

%%%%%%%%%%%%%%%%%%%%%%%%%%%%%%%%%%%%%%%%%%%%%%%%%%%%%%%%%%% 
 \begin{figure}[h!]
 \centering
\begin{tabular}{c}
\includegraphics[width=0.95\linewidth]{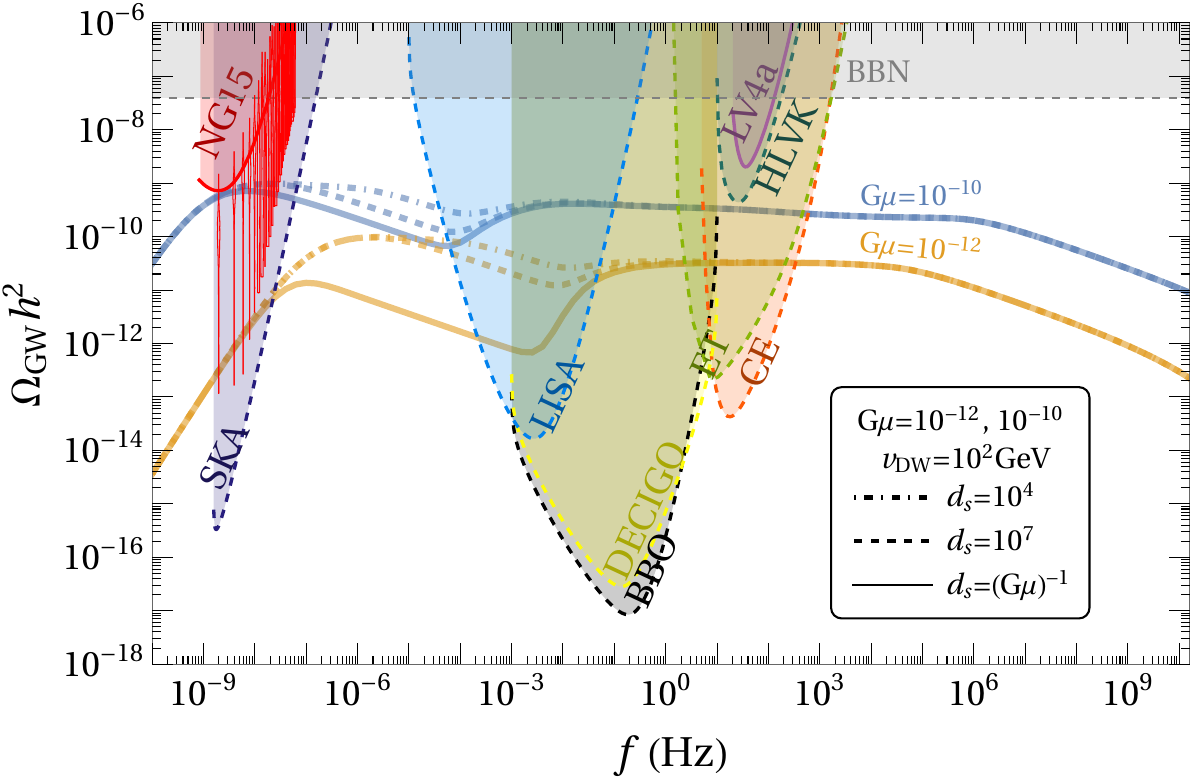} 
 \\
\includegraphics[width=0.95\linewidth]{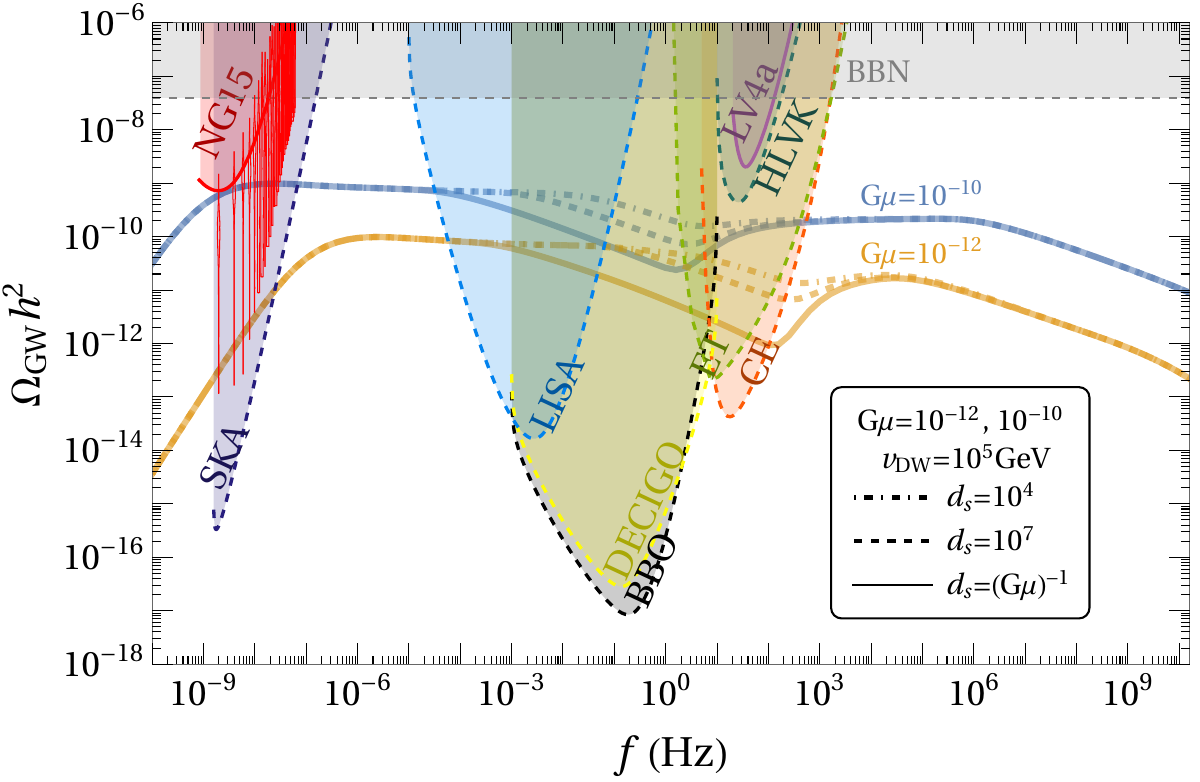} 
\end{tabular}
\caption{Gravitational wave backgrounds from walls bounded by strings and from a network of effectively stable cosmic string network with $G\mu = 10^{-10}$ and $10^{-12}$, and the domain wall VEV $v_{\rm DW}= 10^2$ GeV (upper panel) and $10^5$ GeV (lower panel). We have set the delay factor in Eq.~\eqref{eq:ds} $d_s=10^4, 10^7$ and $(G\mu)^{-1}$. The red violin plots represent the posterior of Hellings-Downs (HD) correlated \cite{Hellings:1983fr} free spectra of NANOGrav 15 year data. We have depicted bounds on the gravitational wave background from big bang nucleosynthesis (BBN) \cite{Maggiore:1999vm}, LVK Run 4a data (LV4a) \cite{LIGOScientific:2025bgj, LIGOScientific:2025kry}, NANOGrav 15 years data (NG15) and the power-law integrated sensitivity curves \cite{Thrane:2013oya, Schmitz:2020syl} for several proposed experiments, namely SKA \cite{Janssen:2014dka}, LISA \cite{Bartolo:2016ami}, DECIGO \cite{Sato_2017}, BBO \cite{Crowder:2005nr, Corbin:2005ny}, HLVK \cite{KAGRA:2013rdx}, ET \cite{Mentasti:2020yyd}, and CE \cite{Regimbau:2016ike}.}
\label{fig:stblstrngplots}
 \end{figure}
%%%%%%%%%%%%%%%%%%%%%%%%%%%%%%%%%%%%%%%%%%%%%%%%%%%%%%%%%%%
 For metastable strings, we express the gravitational wave background as
\begin{align}
\label{eq:gws-mss}
\Omega_{\rm GW}^{\rm str}(f,t_F,R_c) + \Omega_{\rm GW}^{\rm dw}(f)+ \Omega_{\rm GW}^{\rm MSS}(f,t_*,t_{\rm M\bar{M}}),
\end{align}
where $t_*$ is the start time of gravitational waves emission from metastable strings, $t_{\rm M\bar{M}}=1/\sqrt{\Gamma_{\rm M\bar{M}}}$, and $\Gamma_{\rm M\bar{M}}=\tfrac{\mu}{2\pi}\exp(-\pi\kappa)$ represents the decay rate per unit length of the metastable string via quantum tunneling with the metastability factor $\kappa= m_{\rm M}^2/\mu$ ($m_{\rm M}$ is the monopole mass). The gravitational wave background $\Omega_{\rm GW}^{\rm MSS}(f,t_F,t_{\rm M\bar{M}})$ from the metastable strings is estimated as the resultant of the unresolvable bursts from the cusps on the loops. The burst rate is expressed as \cite{Damour:2001bk, Leblond:2009fq, Olmez:2010bi}
\begin{align}
\frac{d^2R_c}{dz \, dl} = N_c \frac{4\pi r^2}{(1+z)^3{H}(z)} \frac{2n(l,t)}{l(1+z)}\Delta_c(f,l,z) ,
\end{align}
where $N_c\sim 2 $ \cite{Cui:2019kkd} is the average number of cusp, $H(z)$ is the Hubble parameter, $r(z)$ represent the proper distance and $\Delta_c(f,l,z) $ is the observable fraction of the  bursts. The loop number distribution is given by \cite{Buchmuller:2021mbb}
\begin{align}
n(l,t)=&
\frac{0.18}{t^{3/2}(l+\Gamma G\mu t)^{5/2}}
\nonumber\\
&\times
\begin{cases}
\Theta(0.18t-l)
\ \ \mathrm{for} ~ t<t_{\rm M\bar{M}},\\[1ex]
\begin{aligned}
&e^{-\Gamma_{\rm M\bar{M}}\left[
l(t-t_{\rm M\bar{M}})
+\frac12\Gamma G\mu
(t-t_{\rm M\bar{M}})^2
\right]}\\
&\times
\Theta(0.18t_{\rm M\bar{M}}-\bar l) \ \ \mathrm{for}~ t>t_{\rm M\bar{M}} ,
\end{aligned}
\end{cases}
\label{eq:source-loop}
\end{align}
with $\bar{l}=l+\Gamma G\mu (t-t_{\rm M\bar{M}})$.
Therefore, the gravitational wave background from the metastable network is given by
\begin{align}
\label{eq:gws-mss2}
\Omega_{\rm GW}^{\rm MSS}(f,t_F,t_{\rm M\bar{M}}) = \frac{4\pi^2}{3H_0^2}f^3\int_{z_{0}}^{z_*} dz \int_0^{d_H} dl \, h^2_c(f,l,z)\frac{d^2R_c}{dz \, dl} \ ,
\end{align}
where $z_*$ denotes the redshift at time $t_*$, $d_H$ is the particle horizon at time $t(z)$,
and $h_c$ denotes the waveform of the burst from a cusp on the loop \cite{Damour:2001bk}.

%%%%%%%%%%%%%%%%%%%%%%%%%%%%%%%%%%%%%%%%%%%%%%%%%%%%%%%%%%%%%%%%%%%%%%%
 \begin{figure}[h!]
\centering
\includegraphics[width=0.95\linewidth]{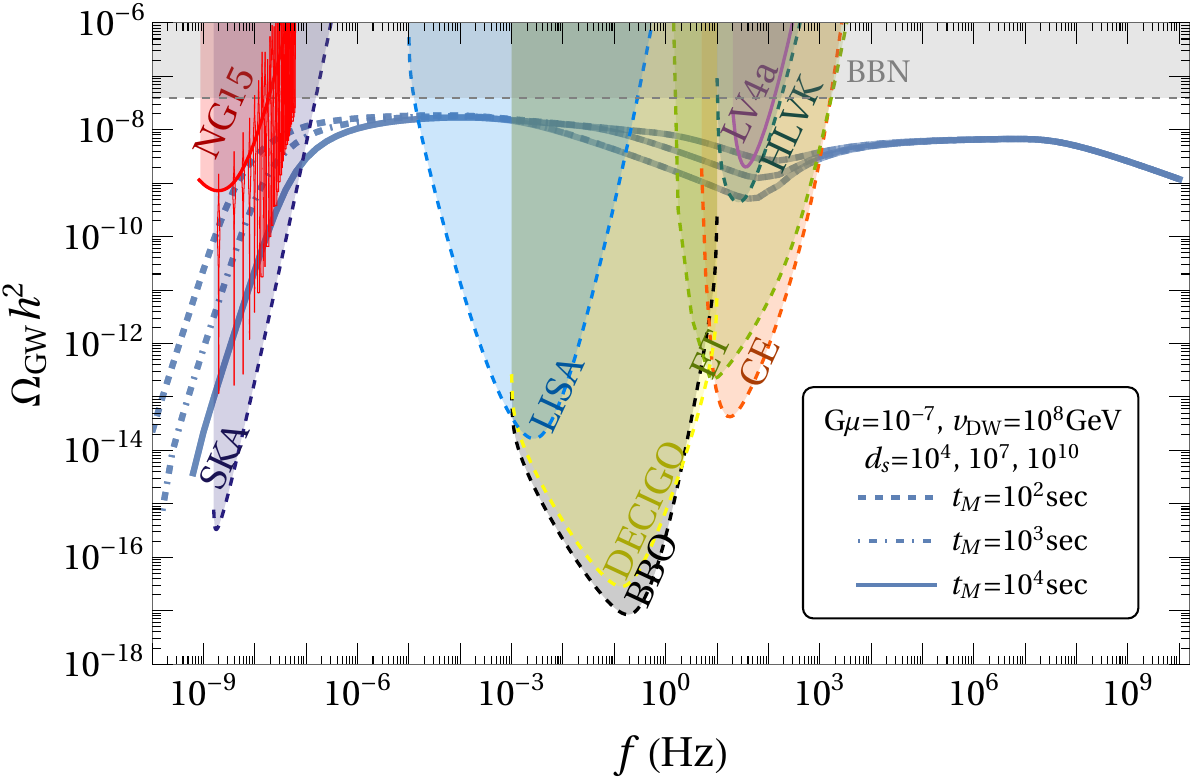}
\caption{ Gravitational wave backgrounds from a quasistable cosmic string network formed from the string-wall composite structures with $G\mu = 10^{-7}$ and $v_{\rm DW}=10^8$~GeV. We set $d_s=10^4,10^7$ and $10^{10}$. For $d_s\gtrsim 10^7$, the spectrum is compatible with the bound from LVK Run 4 data.}
\label{fig:qssplt}
\end{figure}
%%%%%%%%%%%%%%%%%%%%%%%%%%%%%%%%%%%%%%%%%%%%%%%%%%%%%%%%%%%
%%%%%%%%%%%%%%%%%%%%%%%%%%%%%%%%%%%%%%%%%%%%%%%%%%%%%%%%%%% 
 \begin{figure}[h!]
\centering
\includegraphics[width=0.95\linewidth]{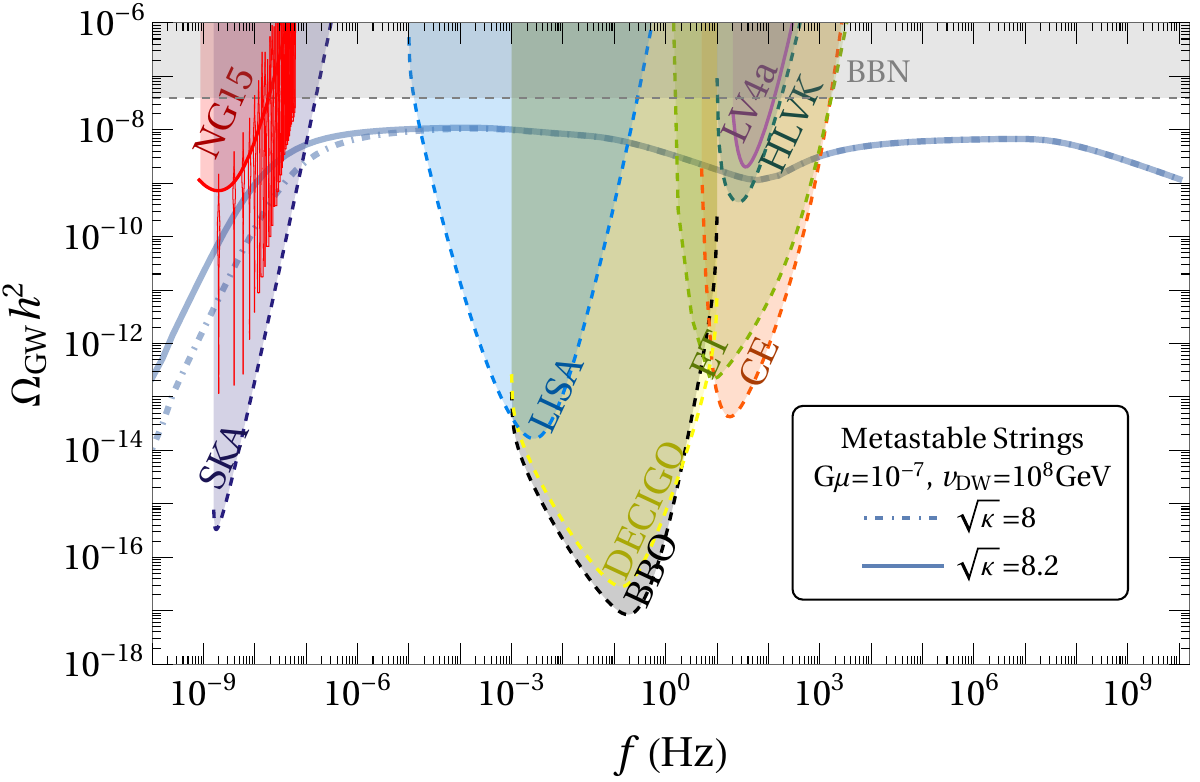}
\caption{Gravitational wave backgrounds from WBS and the metastable cosmic string network formed from the collapsing WBS with $G\mu = 10^{-7}$ and $v_{\rm DW}=10^8$~GeV. The emission of gravitational waves from the metastable strings (see Eqs.~\eqref{eq:gws-mss} and \eqref{eq:gws-mss2}) starts at $t_*=t_d=(G\sigma)^{-1}$. Note that the spectrum is compatible with the recent bound from LVK Run 4a.}
\label{fig:mssplt}
\end{figure}
%%%%%%%%%%%%%%%%%%%%%%%%%%%%%%%%%%%%%%%%%%%%%%%%%%%%%%%%%%%

Fig.~\ref{fig:stblstrngplots} depicts the gravitational wave spectrum for the effectively stable strings from WBS with the primordial string tension $G\mu = 10^{-10}$ and $10^{-12}$, with the domain wall VEV $v_{\rm DW}= 10^2$ GeV and $10^5$ GeV, and we set $d_s=10^4, 10^7$ and $(G\mu)^{-1}$. We have shown the bounds on the gravitational wave background from big bang nucleosynthesis (BBN) \cite{Maggiore:1999vm}, LVK Run 4a data \cite{LIGOScientific:2025bgj, LIGOScientific:2025kry}, and NANOGrav 15 years data (NG15). The red violin plots represent the posterior of
HD-correlated free spectra of NANOGrav 15 year data. We also display the power-law integrated sensitivity curves \cite{Thrane:2013oya, Schmitz:2020syl} for several proposed experiments, namely SKA \cite{Janssen:2014dka}, LISA \cite{Bartolo:2016ami}, DECIGO \cite{Sato_2017}, BBO \cite{Crowder:2005nr, Corbin:2005ny}, HLVK \cite{KAGRA:2013rdx}, ET \cite{Mentasti:2020yyd}, and CE \cite{Regimbau:2016ike}.

 In Fig.~\ref{fig:qssplt}, we show the gravitational wave spectrum from the WBS and the quasistable string network with $G\mu=10^{-7}$ and $t_M = 10^2 - 10^4$ sec.  The spectrum depends on the delay factor $d_s$ around the few Hz frequency range for $v_{\rm DW} = 10^8$ GeV, and for $d_s\gtrsim 10^7$, it is compatible with the LVK Run 4 data. For the WBS and metastable string scenario, we display the gravitational wave spectrum for $G\mu=10^{-7}$, $\sqrt{\kappa}=8-8.2$ and $v_{\rm DW}=10^8$ GeV in Fig.~\ref{fig:mssplt}. The flux matching implies \cite{Maji:2025itv} that monopole-antimonopole nucleation can occur after the merger of two flux tubes bounding a wall at time $t_d$. The start time of gravitational wave emission from metastable strings is taken to be $t_*=t_d$. Similar to the WBS plus quasistable string scenario, the spectrum is compatible with the recent bound from the LVK experiment. 
%%%%%%%%%%%%%%%%%%%%%%%%%%%%%%%%%%%%%%%%%%%%%%%%%%%%%%%%%%%

\section{Conclusions}
\label{sec:conc}
%%%%%%%%%%%%%%%%%%%%%%%%%%%%%%%%%%%%%%%%%%%%%%%%%%%%%%%%%%%
The gauge symmetry breaking $SU(2)
\to U(1) \to Z_2 \to 1$ yields monopoles, strings and finally domain walls bounded by strings. The minimal $SU(2)$ monopole carries a $U(1)$ magnetic flux which is two times larger than the flux carried by the minimal string, namely, ($4 \pi/g$) versus ($2 \pi/g$), where $g$ denotes the $SU(2)$ gauge coupling. However, the collapsing WBS, through mergers of the minimal strings, can yield a network of composite strings that carry the same flux as the minimal monopole. If the primordial monopoles are inflated away, merger of the WBS network can generate a long string network of composite strings carrying two units of $U(1)$ magnetic flux. And, if quantum tunneling through monopole-antimonopole pair creation is suppressed, the composite strings are effectively stable. However, in the presence of quantum tunneling (or if the primordial monopoles reenter the horizon), the string network  is not topologically stable, and we can implement the metastable (or quasistable) string scenario.
We have displayed the stochastic gravitational-wave spectra for these three scenarios which will be tested in the ongoing and future experiments. We briefly show that the same three scenarios are also realized in the symmetry breaking $SU(3) \to SO(3) \to Z_2 \to 1$.

%%%%%%%%%%%%%%%%%%%%%%%%%%%%%%%%%%%%%%%%%%%%%%%%%%%%%%%%%%%%%%%%%%
\section{Acknowledgment}
RM is funded by IBS under the project code: IBS-R018-D3.
%%%%%%%%%%%%%%%%%%%%%%%%%%%%%%%%%%%%%%%%%%%%%%%%%%%%%%%%%%%%%%%%%%%%%%%%%%%%%%%%%%
\appendix

%%%%%%%%%%%%%%%%%%%%%%%%%%
\bibliographystyle{JHEP}
\bibliography{sccs}

\providecommand{\href}[2]{#2}\begingroup\raggedright\begin{thebibliography}{10}

\bibitem{Georgi:1974my}
H.~Georgi, \emph{{The State of the Art\textemdash{}Gauge Theories}},
  \href{https://doi.org/10.1063/1.2947450}{\emph{AIP Conf. Proc.} {\bfseries
  23} (1975) 575}.

\bibitem{Fritzsch:1974nn}
H.~Fritzsch and P.~Minkowski, \emph{{Unified Interactions of Leptons and
  Hadrons}}, \href{https://doi.org/10.1016/0003-4916(75)90211-0}{\emph{Annals
  Phys.} {\bfseries 93} (1975) 193}.

\bibitem{Gursey:1975ki}
F.~Gursey, P.~Ramond and P.~Sikivie, \emph{{A Universal Gauge Theory Model
  Based on E6}},
  \href{https://doi.org/10.1016/0370-2693(76)90417-2}{\emph{Phys. Lett. B}
  {\bfseries 60} (1976) 177}.

\bibitem{Achiman:1978vg}
Y.~Achiman and B.~Stech, \emph{{Quark Lepton Symmetry and Mass Scales in an E6
  Unified Gauge Model}},
  \href{https://doi.org/10.1016/0370-2693(78)90584-1}{\emph{Phys. Lett. B}
  {\bfseries 77} (1978) 389}.

\bibitem{Shafi:1978gg}
Q.~Shafi, \emph{{E(6) as a Unifying Gauge Symmetry}},
  \href{https://doi.org/10.1016/0370-2693(78)90248-4}{\emph{Phys. Lett. B}
  {\bfseries 79} (1978) 301}.

\bibitem{tHooft:1974kcl}
G.~'t~Hooft, \emph{{Magnetic Monopoles in Unified Gauge Theories}},
  \href{https://doi.org/10.1016/0550-3213(74)90486-6}{\emph{Nucl. Phys. B}
  {\bfseries 79} (1974) 276}.

\bibitem{Polyakov:1974ek}
A.M.~Polyakov, \emph{{Particle Spectrum in Quantum Field Theory}}, {\emph{JETP
  Lett.} {\bfseries 20} (1974) 194}.

\bibitem{Lazarides:1980cc}
G.~Lazarides, M.~Magg and Q.~Shafi, \emph{{Phase Transitions and Magnetic
  Monopoles in SO(10)}},
  \href{https://doi.org/10.1016/0370-2693(80)90553-5}{\emph{Phys. Lett. B}
  {\bfseries 97} (1980) 87}.

\bibitem{Kibble:1982ae}
T.W.B.~Kibble, G.~Lazarides and Q.~Shafi, \emph{{Strings in SO(10)}},
  \href{https://doi.org/10.1016/0370-2693(82)90829-2}{\emph{Phys. Lett. B}
  {\bfseries 113} (1982) 237}.

\bibitem{Kibble:1982dd}
T.W.B.~Kibble, G.~Lazarides and Q.~Shafi, \emph{{Walls Bounded by Strings}},
  \href{https://doi.org/10.1103/PhysRevD.26.435}{\emph{Phys. Rev. D} {\bfseries
  26} (1982) 435}.

\bibitem{Makinen:2018ltj}
J.T.~M\"akinen, V.V.~Dmitriev, J.~Nissinen, J.~Rysti, G.E.~Volovik, A.N.~Yudin
  et~al., \emph{{Half-quantum vortices and walls bounded by strings in the
  polar-distorted phases of topological superfluid$^{3}$He}},
  \href{https://doi.org/10.1038/s41467-018-08204-8}{\emph{Nature Commun.}
  {\bfseries 10} (2019) 237}
  [\href{https://arxiv.org/abs/1807.04328}{{\ttfamily 1807.04328}}].

\bibitem{Lazarides:2019xai}
G.~Lazarides and Q.~Shafi, \emph{{Monopoles, Strings, and Necklaces in $SO(10)$
  and $E_6$}}, \href{https://doi.org/10.1007/JHEP10(2019)193}{\emph{JHEP}
  {\bfseries 10} (2019) 193}
  [\href{https://arxiv.org/abs/1904.06880}{{\ttfamily 1904.06880}}].

\bibitem{Lazarides:2023iim}
G.~Lazarides, Q.~Shafi and A.~Tiwari, \emph{{Composite topological structures
  in SO(10)}}, \href{https://doi.org/10.1007/JHEP05(2023)119}{\emph{JHEP}
  {\bfseries 05} (2023) 119}
  [\href{https://arxiv.org/abs/2303.15159}{{\ttfamily 2303.15159}}].

\bibitem{Maji:2025thf}
R.~Maji and Q.~Shafi, \emph{{Superheavy metastable strings in SO(10)}},
  \href{https://doi.org/10.1007/JHEP06(2025)217}{\emph{JHEP} {\bfseries 06}
  (2025) 217} [\href{https://arxiv.org/abs/2504.09055}{{\ttfamily
  2504.09055}}].

\bibitem{LIGOScientific:2016aoc}
{\scshape LIGO Scientific, Virgo} collaboration, \emph{{Observation of
  Gravitational Waves from a Binary Black Hole Merger}},
  \href{https://doi.org/10.1103/PhysRevLett.116.061102}{\emph{Phys. Rev. Lett.}
  {\bfseries 116} (2016) 061102}
  [\href{https://arxiv.org/abs/1602.03837}{{\ttfamily 1602.03837}}].

\bibitem{Buchmuller:2021mbb}
W.~Buchmuller, V.~Domcke and K.~Schmitz, \emph{{Stochastic gravitational-wave
  background from metastable cosmic strings}},
  \href{https://doi.org/10.1088/1475-7516/2021/12/006}{\emph{JCAP} {\bfseries
  12} (2021) 006} [\href{https://arxiv.org/abs/2107.04578}{{\ttfamily
  2107.04578}}].

\bibitem{NANOGrav:2023hvm}
{\scshape NANOGrav} collaboration, \emph{{The NANOGrav 15 yr Data Set: Search
  for Signals from New Physics}},
  \href{https://doi.org/10.3847/2041-8213/acdc91}{\emph{Astrophys. J. Lett.}
  {\bfseries 951} (2023) L11}
  [\href{https://arxiv.org/abs/2306.16219}{{\ttfamily 2306.16219}}].

\bibitem{Lazarides:2022jgr}
G.~Lazarides, R.~Maji and Q.~Shafi, \emph{{Gravitational waves from
  quasi-stable strings}},
  \href{https://doi.org/10.1088/1475-7516/2022/08/042}{\emph{JCAP} {\bfseries
  08} (2022) 042} [\href{https://arxiv.org/abs/2203.11204}{{\ttfamily
  2203.11204}}].

\bibitem{Lazarides:2023ksx}
G.~Lazarides, R.~Maji and Q.~Shafi, \emph{{Superheavy quasistable strings and
  walls bounded by strings in the light of NANOGrav 15~year data}},
  \href{https://doi.org/10.1103/PhysRevD.108.095041}{\emph{Phys. Rev. D}
  {\bfseries 108} (2023) 095041}
  [\href{https://arxiv.org/abs/2306.17788}{{\ttfamily 2306.17788}}].

\bibitem{Maji:2023fba}
R.~Maji, W.-I.~Park and Q.~Shafi, \emph{{Gravitational waves from walls bounded
  by strings in SO(10) model of pseudo-Goldstone dark matter}},
  \href{https://doi.org/10.1016/j.physletb.2023.138127}{\emph{Phys. Lett. B}
  {\bfseries 845} (2023) 138127}
  [\href{https://arxiv.org/abs/2305.11775}{{\ttfamily 2305.11775}}].

\bibitem{NANOGrav:2023gor}
{\scshape NANOGrav} collaboration, \emph{{The NANOGrav 15 yr Data Set: Evidence
  for a Gravitational-wave Background}},
  \href{https://doi.org/10.3847/2041-8213/acdac6}{\emph{Astrophys. J. Lett.}
  {\bfseries 951} (2023) L8}
  [\href{https://arxiv.org/abs/2306.16213}{{\ttfamily 2306.16213}}].

\bibitem{EPTA:2023fyk}
{\scshape EPTA, InPTA:} collaboration, \emph{{The second data release from the
  European Pulsar Timing Array - III. Search for gravitational wave signals}},
  \href{https://doi.org/10.1051/0004-6361/202346844}{\emph{Astron. Astrophys.}
  {\bfseries 678} (2023) A50}
  [\href{https://arxiv.org/abs/2306.16214}{{\ttfamily 2306.16214}}].

\bibitem{Reardon:2023gzh}
D.J.~Reardon et~al., \emph{{Search for an Isotropic Gravitational-wave
  Background with the Parkes Pulsar Timing Array}},
  \href{https://doi.org/10.3847/2041-8213/acdd02}{\emph{Astrophys. J. Lett.}
  {\bfseries 951} (2023) L6}
  [\href{https://arxiv.org/abs/2306.16215}{{\ttfamily 2306.16215}}].

\bibitem{Xu:2023wog}
H.~Xu et~al., \emph{{Searching for the Nano-Hertz Stochastic Gravitational Wave
  Background with the Chinese Pulsar Timing Array Data Release I}},
  \href{https://doi.org/10.1088/1674-4527/acdfa5}{\emph{Res. Astron.
  Astrophys.} {\bfseries 23} (2023) 075024}
  [\href{https://arxiv.org/abs/2306.16216}{{\ttfamily 2306.16216}}].

\bibitem{Dunsky:2021tih}
D.I.~Dunsky, A.~Ghoshal, H.~Murayama, Y.~Sakakihara and G.~White, \emph{{GUTs,
  hybrid topological defects, and gravitational waves}},
  \href{https://doi.org/10.1103/PhysRevD.106.075030}{\emph{Phys. Rev. D}
  {\bfseries 106} (2022) 075030}
  [\href{https://arxiv.org/abs/2111.08750}{{\ttfamily 2111.08750}}].

\bibitem{Maji:2025yms}
R.~Maji and Q.~Shafi, \emph{{C-parity, magnetic monopoles, and higher frequency
  gravitational waves}},
  \href{https://doi.org/10.1103/PhysRevD.111.075027}{\emph{Phys. Rev. D}
  {\bfseries 111} (2025) 075027}
  [\href{https://arxiv.org/abs/2502.10135}{{\ttfamily 2502.10135}}].

\bibitem{Ghoshal:2026wwu}
A.~Ghoshal, I.~Gogoladze and A.~Tiwari, \emph{{Gravitational Waves from hybrid
  defects as probe of Flavor symmetry breaking: Machine-Learning Approach}},
  \href{https://arxiv.org/abs/2605.31600}{{\ttfamily 2605.31600}}.

\bibitem{Antusch:2025xrs}
S.~Antusch, K.~Hinze and S.~Saad, \emph{{Metastable cosmic strings and
  gravitational waves from flavor symmetry breaking}},
  \href{https://doi.org/10.1103/528x-qzs3}{\emph{Phys. Rev. D} {\bfseries 112}
  (2025) 035043} [\href{https://arxiv.org/abs/2503.05868}{{\ttfamily
  2503.05868}}].

\bibitem{Buchmuller:2021dtt}
W.~Buchmuller, \emph{{Metastable strings and dumbbells in supersymmetric hybrid
  inflation}}, \href{https://doi.org/10.1007/JHEP04(2021)168}{\emph{JHEP}
  {\bfseries 04} (2021) 168}
  [\href{https://arxiv.org/abs/2102.08923}{{\ttfamily 2102.08923}}].

\bibitem{Buchmuller:2023aus}
W.~Buchmuller, V.~Domcke and K.~Schmitz, \emph{{Metastable cosmic strings}},
  \href{https://doi.org/10.1088/1475-7516/2023/11/020}{\emph{JCAP} {\bfseries
  11} (2023) 020} [\href{https://arxiv.org/abs/2307.04691}{{\ttfamily
  2307.04691}}].

\bibitem{Lazarides:2023rqf}
G.~Lazarides, R.~Maji, A.~Moursy and Q.~Shafi, \emph{{Inflation, superheavy
  metastable strings and gravitational waves in non-supersymmetric flipped
  SU(5)}}, \href{https://doi.org/10.1088/1475-7516/2024/03/006}{\emph{JCAP}
  {\bfseries 03} (2024) 006}
  [\href{https://arxiv.org/abs/2308.07094}{{\ttfamily 2308.07094}}].

\bibitem{Maji:2023fhv}
R.~Maji and W.-I.~Park, \emph{{Supersymmetric $U(1)_{B-L}$ flat direction and
  NANOGrav 15 year data}},
  \href{https://doi.org/10.1088/1475-7516/2024/01/015}{\emph{JCAP} {\bfseries
  01} (2024) 015} [\href{https://arxiv.org/abs/2308.11439}{{\ttfamily
  2308.11439}}].

\bibitem{Afzal:2023kqs}
A.~Afzal, Q.~Shafi and A.~Tiwari, \emph{{Gravitational wave emission from
  metastable current-carrying strings in E6}},
  \href{https://doi.org/10.1016/j.physletb.2024.138516}{\emph{Phys. Lett. B}
  {\bfseries 850} (2024) 138516}
  [\href{https://arxiv.org/abs/2311.05564}{{\ttfamily 2311.05564}}].

\bibitem{Afzal:2023cyp}
A.~Afzal, M.~Mehmood, M.U.~Rehman and Q.~Shafi, \emph{{Supersymmetric hybrid
  inflation and current-carrying metastable cosmic strings in
  SU(4)c{\texttimes}SU(2)L{\texttimes}U(1)R}},
  \href{https://doi.org/10.1103/d7zx-c7f7}{\emph{Phys. Rev. D} {\bfseries 112}
  (2025) 083545} [\href{https://arxiv.org/abs/2308.11410}{{\ttfamily
  2308.11410}}].

\bibitem{Fu:2023mdu}
B.~Fu, S.F.~King, L.~Marsili, S.~Pascoli, J.~Turner and Y.-L.~Zhou,
  \emph{{Testing realistic SO(10) SUSY GUTs with proton decay and gravitational
  waves}}, \href{https://doi.org/10.1103/PhysRevD.109.055025}{\emph{Phys. Rev.
  D} {\bfseries 109} (2024) 055025}
  [\href{https://arxiv.org/abs/2308.05799}{{\ttfamily 2308.05799}}].

\bibitem{Chitose:2023dam}
A.~Chitose, M.~Ibe, Y.~Nakayama, S.~Shirai and K.~Watanabe, \emph{{Revisiting
  Metastable Cosmic String Breaking}},
  \href{https://arxiv.org/abs/2312.15662}{{\ttfamily 2312.15662}}.

\bibitem{Bao:2024bws}
Y.~Bao, K.~Harigaya and L.-T.~Wang, \emph{{Crescendo beyond the horizon: more
  gravitational waves from domain walls bounded by inflated cosmic strings}},
  \href{https://doi.org/10.1007/JHEP11(2024)032}{\emph{JHEP} {\bfseries 11}
  (2024) 032} [\href{https://arxiv.org/abs/2407.17525}{{\ttfamily
  2407.17525}}].

\bibitem{Maji:2024tzg}
R.~Maji and Q.~Shafi, \emph{{Kinetic mixing, proton decay and gravitational
  waves in SO(10)}}, \href{https://doi.org/10.1007/JHEP10(2024)157}{\emph{JHEP}
  {\bfseries 10} (2024) 157}
  [\href{https://arxiv.org/abs/2408.14350}{{\ttfamily 2408.14350}}].

\bibitem{Maji:2024cwv}
R.~Maji, A.~Moursy and Q.~Shafi, \emph{{Induced gravitational waves, metastable
  cosmic strings and primordial black holes in GUTs}},
  \href{https://doi.org/10.1088/1475-7516/2025/01/106}{\emph{JCAP} {\bfseries
  01} (2025) 106} [\href{https://arxiv.org/abs/2409.13584}{{\ttfamily
  2409.13584}}].

\bibitem{Datta:2024bqp}
S.~Datta and R.~Samanta, \emph{{Cosmic superstrings, metastable strings and
  ultralight primordial black holes: from NANOGrav to LIGO and beyond}},
  \href{https://doi.org/10.1007/JHEP02(2025)095}{\emph{JHEP} {\bfseries 02}
  (2025) 095} [\href{https://arxiv.org/abs/2409.03498}{{\ttfamily
  2409.03498}}].

\bibitem{Chitose:2025qyt}
A.~Chitose, M.~Ibe, S.~Neda and S.~Shirai, \emph{{Do Cosmic String Segments
  Emit Gravitational Waves?}},
  \href{https://arxiv.org/abs/2507.12386}{{\ttfamily 2507.12386}}.

\bibitem{Datta:2025owx}
S.~Datta and R.~Samanta, \emph{{Multifaceted supercooling: from PTA to LIGO}},
  \href{https://doi.org/10.1007/JHEP12(2025)161}{\emph{JHEP} {\bfseries 12}
  (2025) 161} [\href{https://arxiv.org/abs/2506.21397}{{\ttfamily
  2506.21397}}].

\bibitem{Ingoldby:2025wcl}
J.~Ingoldby, V.V.~Khoze and J.~Turner, \emph{{Metastable strings and
  gravitational waves in one-scale models}},
  \href{https://doi.org/10.1007/JHEP04(2026)094}{\emph{JHEP} {\bfseries 04}
  (2026) 094} [\href{https://arxiv.org/abs/2511.08546}{{\ttfamily
  2511.08546}}].

\bibitem{Tranchedone:2026lav}
L.~Tranchedone, E.~Carragher, E.~Hardy and N.K.~van IJcken, \emph{{Metastable
  cosmic strings are broken at the start}},
  \href{https://arxiv.org/abs/2601.04320}{{\ttfamily 2601.04320}}.

\bibitem{Asl:2026zpj}
D.H.~Asl and K.~Schmitz, \emph{{New gravitational-wave templates for metastable
  cosmic strings: Loop breaking versus network collapse}},
  \href{https://arxiv.org/abs/2604.28097}{{\ttfamily 2604.28097}}.

\bibitem{Blasi:2026iyq}
S.~Blasi, M.~Grandjean and A.~Mariotti, \emph{{Metastable strings at PTAs:
  classical stability analysis}},
  \href{https://arxiv.org/abs/2605.03003}{{\ttfamily 2605.03003}}.

\bibitem{deGiorgi:2026fyx}
A.~de~Giorgi, J.~Ingoldby, V.V.~Khoze and J.~Turner, \emph{{Thermal Metastable
  Strings in One-Scale Models and Gravitational Waves}},
  \href{https://arxiv.org/abs/2606.02689}{{\ttfamily 2606.02689}}.

\bibitem{Hua:2026mgn}
J.~Hua, B.~Fu and Y.-L.~Tang, \emph{{One-Dimensional Simulations of the
  Topological Defects in a 3:1 $U(1)$ Model}},
  \href{https://arxiv.org/abs/2607.13066}{{\ttfamily 2607.13066}}.

\bibitem{LIGOScientific:2025bgj}
{\scshape LIGO Scientific, VIRGO, KAGRA} collaboration, \emph{{Upper Limits on
  the Isotropic Gravitational-Wave Background from the first part of LIGO,
  Virgo, and KAGRA's fourth Observing Run}},
  \href{https://arxiv.org/abs/2508.20721}{{\ttfamily 2508.20721}}.

\bibitem{LIGOScientific:2025kry}
{\scshape LIGO Scientific, VIRGO, KAGRA} collaboration, \emph{{Cosmological and
  High Energy Physics implications from gravitational-wave background searches
  in LIGO-Virgo-KAGRA's O1-O4a runs}},
  \href{https://arxiv.org/abs/2510.26848}{{\ttfamily 2510.26848}}.

\bibitem{Vachaspati:1984gt}
T.~Vachaspati and A.~Vilenkin, \emph{{Gravitational Radiation from Cosmic
  Strings}}, \href{https://doi.org/10.1103/PhysRevD.31.3052}{\emph{Phys. Rev.
  D} {\bfseries 31} (1985) 3052}.

\bibitem{Vilenkin:2000jqa}
A.~Vilenkin and E.P.S.~Shellard, \emph{{Cosmic Strings and Other Topological
  Defects}}, Cambridge University Press (7, 2000).

\bibitem{Hiramatsu:2013qaa}
T.~Hiramatsu, M.~Kawasaki and K.~Saikawa, \emph{{On the estimation of
  gravitational wave spectrum from cosmic domain walls}},
  \href{https://doi.org/10.1088/1475-7516/2014/02/031}{\emph{JCAP} {\bfseries
  02} (2014) 031} [\href{https://arxiv.org/abs/1309.5001}{{\ttfamily
  1309.5001}}].

\bibitem{Martin:1996cp}
X.~Martin and A.~Vilenkin, \emph{{Gravitational radiation from monopoles
  connected by strings}},
  \href{https://doi.org/10.1103/PhysRevD.55.6054}{\emph{Phys. Rev. D}
  {\bfseries 55} (1997) 6054}
  [\href{https://arxiv.org/abs/gr-qc/9612008}{{\ttfamily gr-qc/9612008}}].

\bibitem{Dvali:2022vwh}
G.~Dvali, J.S.~Valbuena-Berm{\'u}dez and M.~Zantedeschi, \emph{{Dynamics of
  confined monopoles and similarities with confined quarks}},
  \href{https://doi.org/10.1103/PhysRevD.107.076003}{\emph{Phys. Rev. D}
  {\bfseries 107} (2023) 076003}
  [\href{https://arxiv.org/abs/2210.14947}{{\ttfamily 2210.14947}}].

\bibitem{Vanchurin:2005pa}
V.~Vanchurin, K.D.~Olum and A.~Vilenkin, \emph{{Scaling of cosmic string
  loops}}, \href{https://doi.org/10.1103/PhysRevD.74.063527}{\emph{Phys. Rev.
  D} {\bfseries 74} (2006) 063527}
  [\href{https://arxiv.org/abs/gr-qc/0511159}{{\ttfamily gr-qc/0511159}}].

\bibitem{Ringeval:2005kr}
C.~Ringeval, M.~Sakellariadou and F.~Bouchet, \emph{{Cosmological evolution of
  cosmic string loops}},
  \href{https://doi.org/10.1088/1475-7516/2007/02/023}{\emph{JCAP} {\bfseries
  02} (2007) 023} [\href{https://arxiv.org/abs/astro-ph/0511646}{{\ttfamily
  astro-ph/0511646}}].

\bibitem{Olum:2006ix}
K.D.~Olum and V.~Vanchurin, \emph{{Cosmic string loops in the expanding
  Universe}}, \href{https://doi.org/10.1103/PhysRevD.75.063521}{\emph{Phys.
  Rev. D} {\bfseries 75} (2007) 063521}
  [\href{https://arxiv.org/abs/astro-ph/0610419}{{\ttfamily
  astro-ph/0610419}}].

\bibitem{Olmez:2010bi}
S.~Olmez, V.~Mandic and X.~Siemens, \emph{{Gravitational-Wave Stochastic
  Background from Kinks and Cusps on Cosmic Strings}},
  \href{https://doi.org/10.1103/PhysRevD.81.104028}{\emph{Phys. Rev. D}
  {\bfseries 81} (2010) 104028}
  [\href{https://arxiv.org/abs/1004.0890}{{\ttfamily 1004.0890}}].

\bibitem{Blanco-Pillado:2013qja}
J.J.~Blanco-Pillado, K.D.~Olum and B.~Shlaer, \emph{{The number of cosmic
  string loops}}, \href{https://doi.org/10.1103/PhysRevD.89.023512}{\emph{Phys.
  Rev. D} {\bfseries 89} (2014) 023512}
  [\href{https://arxiv.org/abs/1309.6637}{{\ttfamily 1309.6637}}].

\bibitem{Blanco-Pillado:2017oxo}
J.J.~Blanco-Pillado and K.D.~Olum, \emph{{Stochastic gravitational wave
  background from smoothed cosmic string loops}},
  \href{https://doi.org/10.1103/PhysRevD.96.104046}{\emph{Phys. Rev. D}
  {\bfseries 96} (2017) 104046}
  [\href{https://arxiv.org/abs/1709.02693}{{\ttfamily 1709.02693}}].

\bibitem{Cui:2018rwi}
Y.~Cui, M.~Lewicki, D.E.~Morrissey and J.D.~Wells, \emph{{Probing the pre-BBN
  universe with gravitational waves from cosmic strings}},
  \href{https://doi.org/10.1007/JHEP01(2019)081}{\emph{JHEP} {\bfseries 01}
  (2019) 081} [\href{https://arxiv.org/abs/1808.08968}{{\ttfamily
  1808.08968}}].

\bibitem{Damour:2001bk}
T.~Damour and A.~Vilenkin, \emph{{Gravitational wave bursts from cusps and
  kinks on cosmic strings}},
  \href{https://doi.org/10.1103/PhysRevD.64.064008}{\emph{Phys. Rev. D}
  {\bfseries 64} (2001) 064008}
  [\href{https://arxiv.org/abs/gr-qc/0104026}{{\ttfamily gr-qc/0104026}}].

\bibitem{Martin:1996ea}
X.~Martin and A.~Vilenkin, \emph{{Gravitational wave background from hybrid
  topological defects}},
  \href{https://doi.org/10.1103/PhysRevLett.77.2879}{\emph{Phys. Rev. Lett.}
  {\bfseries 77} (1996) 2879}
  [\href{https://arxiv.org/abs/astro-ph/9606022}{{\ttfamily
  astro-ph/9606022}}].

\bibitem{Planck:2018vyg}
{\scshape Planck} collaboration, \emph{{Planck 2018 results. VI. Cosmological
  parameters}},
  \href{https://doi.org/10.1051/0004-6361/201833910}{\emph{Astron. Astrophys.}
  {\bfseries 641} (2020) A6}
  [\href{https://arxiv.org/abs/1807.06209}{{\ttfamily 1807.06209}}] [Erratum:
  \href{https://www.aanda.org/10.1051/0004-6361/201833910e}{$Astron.~Astrophys.$
  {\bf 652} (2021) C4]}.

\bibitem{Vilenkin:1991zk}
A.~Vilenkin, \emph{{Cosmic string dynamics with friction}},
  \href{https://doi.org/10.1103/PhysRevD.43.1060}{\emph{Phys. Rev. D}
  {\bfseries 43} (1991) 1060}.

\bibitem{Garriga:1993gj}
J.~Garriga and M.~Sakellariadou, \emph{{Effects of friction on cosmic
  strings}}, \href{https://doi.org/10.1103/PhysRevD.48.2502}{\emph{Phys. Rev.
  D} {\bfseries 48} (1993) 2502}
  [\href{https://arxiv.org/abs/hep-th/9303024}{{\ttfamily hep-th/9303024}}].

\bibitem{Blanco-Pillado:1998tyu}
J.J.~Blanco-Pillado and K.D.~Olum, \emph{{Form of cosmic string cusps}},
  \href{https://doi.org/10.1103/PhysRevD.59.063508}{\emph{Phys. Rev. D}
  {\bfseries 59} (1999) 063508}
  [\href{https://arxiv.org/abs/gr-qc/9810005}{{\ttfamily gr-qc/9810005}}]
  [Erratum: Phys.Rev.D 103, 029902 (2021)].

\bibitem{Matsunami:2019fss}
D.~Matsunami, L.~Pogosian, A.~Saurabh and T.~Vachaspati, \emph{{Decay of Cosmic
  String Loops Due to Particle Radiation}},
  \href{https://doi.org/10.1103/PhysRevLett.122.201301}{\emph{Phys. Rev. Lett.}
  {\bfseries 122} (2019) 201301}
  [\href{https://arxiv.org/abs/1903.05102}{{\ttfamily 1903.05102}}].

\bibitem{Auclair:2019jip}
P.~Auclair, D.A.~Steer and T.~Vachaspati, \emph{{Particle emission and
  gravitational radiation from cosmic strings: observational constraints}},
  \href{https://doi.org/10.1103/PhysRevD.101.083511}{\emph{Phys. Rev. D}
  {\bfseries 101} (2020) 083511}
  [\href{https://arxiv.org/abs/1911.12066}{{\ttfamily 1911.12066}}].

\bibitem{Hellings:1983fr}
R.W.~Hellings and G.S.~Downs, \emph{{Upper limits on the isotropic
  gravitational radiation background from pulsar timing analysis}},
  \href{https://doi.org/10.1086/183954}{\emph{Astrophys. J. Lett.} {\bfseries
  265} (1983) L39}.

\bibitem{Maggiore:1999vm}
M.~Maggiore, \emph{{Gravitational wave experiments and early universe
  cosmology}}, \href{https://doi.org/10.1016/S0370-1573(99)00102-7}{\emph{Phys.
  Rept.} {\bfseries 331} (2000) 283}
  [\href{https://arxiv.org/abs/gr-qc/9909001}{{\ttfamily gr-qc/9909001}}].

\bibitem{Thrane:2013oya}
E.~Thrane and J.D.~Romano, \emph{{Sensitivity curves for searches for
  gravitational-wave backgrounds}},
  \href{https://doi.org/10.1103/PhysRevD.88.124032}{\emph{Phys. Rev. D}
  {\bfseries 88} (2013) 124032}
  [\href{https://arxiv.org/abs/1310.5300}{{\ttfamily 1310.5300}}].

\bibitem{Schmitz:2020syl}
K.~Schmitz, \emph{{New Sensitivity Curves for Gravitational-Wave Signals from
  Cosmological Phase Transitions}},
  \href{https://doi.org/10.1007/JHEP01(2021)097}{\emph{JHEP} {\bfseries 01}
  (2021) 097} [\href{https://arxiv.org/abs/2002.04615}{{\ttfamily
  2002.04615}}].

\bibitem{Janssen:2014dka}
G.~Janssen et~al., \emph{{Gravitational wave astronomy with the SKA}},
  \href{https://doi.org/10.22323/1.215.0037}{\emph{PoS} {\bfseries AASKA14}
  (2015) 037} [\href{https://arxiv.org/abs/1501.00127}{{\ttfamily
  1501.00127}}].

\bibitem{Bartolo:2016ami}
N.~Bartolo et~al., \emph{{Science with the space-based interferometer LISA. IV:
  Probing inflation with gravitational waves}},
  \href{https://doi.org/10.1088/1475-7516/2016/12/026}{\emph{JCAP} {\bfseries
  12} (2016) 026} [\href{https://arxiv.org/abs/1610.06481}{{\ttfamily
  1610.06481}}].

\bibitem{Sato_2017}
S.~Sato et~al., \emph{The status of {DECIGO}},
  \href{https://doi.org/10.1088/1742-6596/840/1/012010}{\emph{Journal of
  Physics: Conference Series} {\bfseries 840} (2017) 012010}.

\bibitem{Crowder:2005nr}
J.~Crowder and N.J.~Cornish, \emph{{Beyond LISA: Exploring future gravitational
  wave missions}},
  \href{https://doi.org/10.1103/PhysRevD.72.083005}{\emph{Phys. Rev. D}
  {\bfseries 72} (2005) 083005}
  [\href{https://arxiv.org/abs/gr-qc/0506015}{{\ttfamily gr-qc/0506015}}].

\bibitem{Corbin:2005ny}
V.~Corbin and N.J.~Cornish, \emph{{Detecting the cosmic gravitational wave
  background with the big bang observer}},
  \href{https://doi.org/10.1088/0264-9381/23/7/014}{\emph{Class. Quant. Grav.}
  {\bfseries 23} (2006) 2435}
  [\href{https://arxiv.org/abs/gr-qc/0512039}{{\ttfamily gr-qc/0512039}}].

\bibitem{KAGRA:2013rdx}
{\scshape KAGRA, LIGO Scientific, Virgo, VIRGO} collaboration, \emph{{Prospects
  for observing and localizing gravitational-wave transients with Advanced
  LIGO, Advanced Virgo and KAGRA}},
  \href{https://doi.org/10.1007/s41114-020-00026-9}{\emph{Living Rev. Rel.}
  {\bfseries 21} (2018) 3} [\href{https://arxiv.org/abs/1304.0670}{{\ttfamily
  1304.0670}}].

\bibitem{Mentasti:2020yyd}
G.~Mentasti and M.~Peloso, \emph{{ET sensitivity to the anisotropic Stochastic
  Gravitational Wave Background}},
  \href{https://doi.org/10.1088/1475-7516/2021/03/080}{\emph{JCAP} {\bfseries
  03} (2021) 080} [\href{https://arxiv.org/abs/2010.00486}{{\ttfamily
  2010.00486}}].

\bibitem{Regimbau:2016ike}
T.~Regimbau, M.~Evans, N.~Christensen, E.~Katsavounidis, B.~Sathyaprakash and
  S.~Vitale, \emph{{Digging deeper: Observing primordial gravitational waves
  below the binary black hole produced stochastic background}},
  \href{https://doi.org/10.1103/PhysRevLett.118.151105}{\emph{Phys. Rev. Lett.}
  {\bfseries 118} (2017) 151105}
  [\href{https://arxiv.org/abs/1611.08943}{{\ttfamily 1611.08943}}].

\bibitem{Leblond:2009fq}
L.~Leblond, B.~Shlaer and X.~Siemens, \emph{{Gravitational Waves from Broken
  Cosmic Strings: The Bursts and the Beads}},
  \href{https://doi.org/10.1103/PhysRevD.79.123519}{\emph{Phys. Rev. D}
  {\bfseries 79} (2009) 123519}
  [\href{https://arxiv.org/abs/0903.4686}{{\ttfamily 0903.4686}}].

\bibitem{Cui:2019kkd}
Y.~Cui, M.~Lewicki and D.E.~Morrissey, \emph{{Gravitational Wave Bursts as
  Harbingers of Cosmic Strings Diluted by Inflation}},
  \href{https://doi.org/10.1103/PhysRevLett.125.211302}{\emph{Phys. Rev. Lett.}
  {\bfseries 125} (2020) 211302}
  [\href{https://arxiv.org/abs/1912.08832}{{\ttfamily 1912.08832}}].

\bibitem{Maji:2025itv}
R.~Maji and Q.~Shafi, \emph{{Superconducting strings in $E_6$}},
  \href{https://doi.org/10.1103/nym6-vpms}{\emph{Phys. Rev. D} {\bfseries 112}
  (2025) L101903} [\href{https://arxiv.org/abs/2509.15985}{{\ttfamily
  2509.15985}}].

\end{thebibliography}\endgroup
%%%%%%%%%%%%%%%%%%%%%%%%%%

\end{document}